\documentclass[journal,article,submit,moreauthors,pdftex,10pt,a4paper]{mdpi}

\usepackage{amsmath,amssymb,amstext,graphicx,bm,bbold,multirow,color,gensymb}

\usepackage[utf8]{inputenc}

\DeclareMathOperator\Tr{Tr}

\firstpage{1} 
\articlenumber{x}
\doinum{10.3390/------}
\pubvolume{xx}
\pubyear{2017}
\copyrightyear{2017}
\externaleditor{Academic Editor: Jordi Cirera}
\history{Received: date; Accepted: date; Published: date}

\Title{Magnetic properties of metal-organic coordination networks based  on 3d transition metal atoms }

\Author{Mar\'{\i}a Blanco-Rey $^{1,2}$, Ane Sarasola $^{2,3}$, Corneliu Nistor $^{4}$, Luca Persichetti $^{4}$, Christian Stamm $^{4}$, Cinthia Piamonteze $^{5}$, Pietro Gambardella $^{4}$, Sebastian Stepanow $^{4}$, Mikhail M. Otrokov $^{2,6,8,9}$, Vitaly N. Golovach  $^{1,2,6,7}$ and Andres Arnau $^{1,2,6}$}

\address{%

$^{1}$ \quad Departamento de F\'{\i}sica de Materiales UPV/EHU, 20018 Donostia-San Sebasti\'{a}n, Spain\\
$^{2}$ \quad Donostia International Physics Center (DIPC),  20018 Donostia-San Sebasti\'{a}n, Spain\\
$^{3}$ \quad Departamento F\'{\i}sica Aplicada I, Universidad del Pa\'{\i}s Vasco, 20018 Donostia-San Sebasti\'{a}n, Spain\\
$^{4}$ \quad Department of Materials, ETH Z\"urich, H\"onggerbergring 64, 8093 Z\"urich, Switzerland\\
$^{5}$ \quad Swiss Light Source, Paul Scherrer Institute, 5232 Villigen PSI, Switzerland\\
$^{6}$ \quad Centro de F\'{i}sica de Materiales (CFM-MPC), Centro Mixto CSIC-UPV/EHU,  20018 Donostia-San Sebasti\'{a}n, Basque Country, Spain\\
$^{7}$ \quad IKERBASQUE, Basque Foundation for Science, 48013 Bilbao, Basque Country, Spain\\
$^{8}$ \quad Tomsk State University, 634050 Tomsk, Russia\\
$^{9}$ \quad Saint Petersburg State University, 198504 Saint Petersburg, Russia}

\abstract{The magnetic anisotropy and exchange coupling between spins localized at the positions of 3d transition metal atoms forming two-dimensional metal-organic coordination networks (MOCNs) grown on the Au(111) metal surface  are studied. 
In particular, we consider MOCNs made of Ni or Mn metal centers linked by TCNQ (7,7,8,8-tetracyanoquinodimethane) organic ligands, which form rectangular networks with 1:1 stoichiometry. Based on the analysis of X-ray magnetic circular dichroism (XMCD) data taken at T= 2.5 K, we find that Ni  atoms in the Ni-TCNQ MOCNs are coupled ferromagnetically and do not show any significant magnetic anisotropy, while Mn atoms in the Mn-TCNQ MOCNs are coupled antiferromagnetically and do show a weak magnetic anisotropy with in-plane magnetization. We explain these observations using both a model Hamiltonian based on mean-field Weiss theory and density functional theory calculations that include spin-orbit coupling. Our main conclusion is that the antiferromagnetic coupling between Mn spins and the in-plane magnetization of the Mn spins can be explained neglecting effects due to the presence of the Au(111) surface, while for Ni-TCNQ the metal surface plays a role in determining the absence of magnetic anisotropy in the system.}

\keyword{magnetism; metal-organic network; XMCD; Density Functional Theory}

\begin{document}


\section{Introduction}
\label{intro}

There exists an exciting type of two-dimensional systems that can be grown on surfaces by self-assembly techniques with interest both from a fundamental point of view and also due to its potential application in the fabrication of electronic and spintronic devices. They are called metal-organic coordination networks (MOCNs) and consist of metal centers linked by organic ligands that permit, in principle, the design of overlayers with specific electronic and magnetic properties \cite{DONG2016101}. 
The synthesis and growth of a given MOCN with a given composition, essentially defined by its stoichiometry and coordination, depends on the relative strength of the interactions between the constituents (organic ligands and metal centers) and their interaction with the underlying surface \cite{Wegner2008,Bedwani2008,Faraggi2012,Umbach2012,Abdurakhmanova2012,Abdurakhmanova2013,Giovanelli2014,Bebensee2014,Otero2015,Faraggi2015}. Indeed, the chemical state of the organic ligands and metal centers can be modified due to vertical electronic charge transfer from the surface  \cite{OTERO2017105}. Additionally, lateral charge transfer between the MOCN's constituents is crucial for the bonding and equally important for the electronic and chemical properties of the overlayers. Particularly interesting is the role of the metal centers in the formation of the two-dimensional networks  by favoring a given coordination and stoichiometry determining the charge and magnetic moment of the metal center and, occasionally, also of the organic ligand that can acquire spin polarization. The important point is that this very fact could be used to control the electronic and magnetic properties of the interface.

The case of 3d transition metal atoms as metal centers and molecules with large electronegativity, like TCNQ (7,7,8,8-tetracyanoquinodimethane) or F4TCNQ (2,3,5,6-tetrafluoro-7,7,8,8-tetracyanoquinodimethane), on metal surfaces is of special interest because they form well-ordered MOCNs with few defects \cite{Tseng2010, Faraggi2012,Abdurakhmanova2013} and different stoichiometry; this latter depending both on the underlying surface 
and preparation conditions. The experimental techniques typically used to characterize the geometric structure and chemical composition of MOCNs on surfaces are scanning tunneling microscopy (STM), low-energy electron difraction (LEED) and X-ray photoemission spectroscopy (XPS), while for the electronic and magnetic properties the standard techniques are angle resolved photoemission spectroscopy (ARPES) , scanning tunneling spectroscopy (STS) and X-ray magnetic circular dichroism (XMCD).

The theoretical description of this type of systems has been shown to be quite reasonable using standard electronic structure methods, like density functional theory (DFT) \cite{Tseng2010, Faraggi2012, Faraggi2015, PhysRevB.92.184424}, although one has to be aware of the limitations of each method when aiming at achieving quantitative agreement with the experimental data. However, the explanation of the observations at a qualitative level and its understanding with the help of a model Hamiltonian is the recipe that we follow in this work. In any case, it is worth to mention that an essential problem, which is hard to overcome, concerns the accuracy of the calculations when dealing with very low energy scales, as it is the case of the determination of exchange couplings or magnetic anisotropies in the sub-meV energy range. Apart from this limitation imposed by the methodology itself, it is also important  to stress that exchange coupling and magnetic anisotropy energy are extremely sensitive to both slight geometrical distortions or band filling, i. e., electronic charge transfer. Therefore, it is important to balance the different effects that each and every approximation can have in the final results when trying to explain an observation that can be deduced, e. g., from XMCD data, regarding the strength and type (ferro- or antiferro-magnetic) of exchange coupling or the kind of magnetic anisotropy (easy axis or easy plane). 

The exchange coupling between magnetic centers in two-dimensional MOCNs is affected to a major or lesser extent by the underlying substrate. The presence of the surface represents a difficulty to describe the full system (MOCN/surface) because the overlayer structure is not necessarily commensurable with the crystal surface or because the size of the commensurable supercell is too large. 
In the case of weak coupling between the overlayer and the surface, as it is the case of Au(111) surfaces, the essential features can be described by neglecting the role of surface electrons, in first approximation. As a rule of thumb, when the lateral bonds between the metal centers and the organic ligands are much stronger than the metal-surface or ligand-surface bonds, this approximation is expected to be a reasonable way to describe the magnetic coupling between metal atoms in the MOCN. However, in case of lateral (metal-molecule or intermolecular) and vertical (MOCN-surface) couplings of comparable strengths, an explicit inclusion of the surface is required to describe the system, something that could happen either due to strong coupling with the surface or weak lateral coupling. 
Next, one should consider the role of charge transfer between surface and overlayer, even in the case of weak coupling, as it can be important in determining magnetic moments, magnetic coupling or even magnetic anisotropy. Indeed, when intermolecular coupling is weak, the role of surface electrons can be relatively more important in determining the magnetic coupling between spins of the metal centers via RKKY interaction \cite{Yosida1996}, which may appear not only on metal surfaces but also on the surface of topological insulators \cite{PhysRevLett.106.136802,Caputo.nl2016}. Very recently, it has been proposed that the RKKY interaction is responsible for the long range ferrimagnetic order \cite{Girovsky2017} in a two-dimensional Kondo lattice with underscreened spins by the conduction electrons in a FeFPc - MnPc mixture on the Au(111) surface.

In this work, we consider the case of MOCNs that consist of Mn or Ni magnetic atoms and TCNQ or F4TCNQ molecules grown on Au(111) surfaces, which show 1:1 stoichiometry with each metal center (Mn or Ni) coordinated with four organic ligands. 
For these systems, our XMCD data show a substantial difference between Mn and Ni networks both for TCNQ and F4TCNQ ligands that we explain using a model Hamiltonian approach and DFT calculations.
The results of our calculations for the free-standing neutral Mn-TCNQ overlayers are consistent with both the antiferromagnetic (AFM) coupling between Mn centers and the weak magnetic anisotropy with in-plane magnetization, while for Ni-TCNQ overlayers we need to call for effects due to the presence of the underlying metal surface, like charge transfer and changes in coordination,  to explain the absence of anisotropy in the system. 
Model calculations based on mean-field Weiss theory permit us to extract exchange coupling constants from the fits to XMCD curves, as well as to obtain additional information about the magnetic anisotropy and the different magnetic configurations that may appear in the networks. Finally, it is worth mentioning that, although the organic ligands TCNQ and F4TCNQ have a different electronegativity (higher in F4TCNQ than in TCNQ), based on the acquired XMCD data, there are not substantial differences in the magnetic properties of the corresponding Ni and Mn networks. Therefore, in the core of the paper we present the results for TCNQ networks and leave the F4TCNQ results for Section I of the Supplementary Material. 

The paper is organized as follows. After describing the XMCD experiments and the technical details of the calculations in section~\ref{MandM}, we present  our XMCD data for Mn-TCNQ and Ni-TCNQ on Au(111) in Section~\ref{XMCD} , together with fitting curves from model calculations that permit to explain the observations and extract information about the type of magnetic coupling and magnetic anisotropy (Section ~\ref{modelH}). Next, in Section ~\ref{DFT} we present the results of our spin polarized DFT+U electronic structure calculations for Mn-TCNQ and Ni-TCNQ free-standing overlayers that confirm the observed behavior in the type of magnetic coupling between spins at the 3d metal centers.
 Then, in Section ~\ref{MAE}, we present the magneto-crystalline anisotropy analysis of the two considered systems under study based on calculations that include spin-orbit coupling (SOC). Finally, in Section ~\ref{DandC}, we present a discussion of our findings and establish the main conclusions that aim at explaining the XMCD observations and suggest that for Ni-TCNQ networks the Au(111) metal surface plays a role in determining the magnetic properties of the MOCN, while this is not the case for Mn-TCNQ.

 

\section{Materials and Methods}
\label{MandM}

\subsection{XMCD experiments}

The x-ray absorption spectroscopy (XAS) experiments were carried out at the X-Treme beamline of the Swiss Light Source. The samples were prepared in ultra-high vacuum chambers with a base pressure in the range of low 10$^{-10}$~mbar. The pressure in the magnet-cryo-chamber was always better than 10$^{-11}$~mbar. The Au(111) surface was cleaned by repeated cycles of Ar$^+$ sputtering and subsequent annealing to 800~K. The molecules TCNQ (7,7,8,8-tetracyanoquinodimethane, 98\% purity, Aldrich) and F4TCNQ (2,3,5,6-tetrafluoro-7,7,8,8-tetracyanoquinodimethane, 97\% purity, Aldrich) were thoroughly degassed before evaporation. The organic adlayers were grown by organic molecular-beam epitaxy (OMBE) using a resistively heated quartz crucible at a sublimation temperature of 408~K onto the clean Au(111) surface that was kept at room temperature. The coverage of molecules was controlled to be below one monolayer. Ni or Mn was subsequently deposited using an electron-beam heating evaporator at a flux of about 0.01~ML/min on top of the molecular adlayers that were heated to 350-400~K to promote the network formation. The sample was checked in-situ by STM at the beamline and subsequently transferred to magnet chamber without breaking the vacuum.

The polarization-dependent XAS experiments were performed in total electron yield detection. Magnetic fields were applied collinear with the photon beam at sample temperatures between 2.5 and 300~K. The data were acquired by varying the photon energy at the L$_{2,3}$~edges of Ni and Mn, as well as the  K~edges of O and N using circular and linear polarized light. The absorption spectra were normalized with respect to the total flux of the incoming x-rays and were further treated to be normalized to the absorption pre-edge due to TEY variations. The background obtained from clean or molecule-covered Au(111) was subtracted to allow comparison of the spectral features. The XMCD is obtained from the difference of the left and right circular polarized XAS spectra whereas the XAS is obtained from the average of the two circular polarizations. The sample was rotated between normal x-ray incidence with respect to the sample surface at $\theta = 0^\circ$ and grazing incidence with $\theta = 60^\circ$. All shown spectra were acquired at $T = 2.5$~K at external magnetic fields up to $\mu_0 H=6.8$~T. The magnetization curves were recorded by acquiring the maximum of the XMCD signal at the L$_3$~ edge as a function of the external magnetic field, normalized by the corresponding pre-edge of the XAS signal. To facilitate the extraction of the easy and hard magnetization axis the magnetization curves at different angles of the magnetic field were normalized to the same value at the highest magnetic field point.

\subsection{Density Functional Theory calculations}

DFT calculations have been carried out 
using the Vienna Ab Initio Simulation Package (VASP) \cite{Kresse93,Kresse96,Kresse96b}. For the description of 
electron-ion interactions the Projector Augmented-Wave (PAW) method has been employed, whereas  the Perdew, Burke and
Ernzerhof (PBE) \cite{PBE96} functional has been used to describe exchange and correlation within the generalized gradient approximation (GGA). 
A Hubbard-like Coulomb repulsion correction term (U= 4 eV) has been added to describe 
the 3d metal electron states, based on Dudarev's approach \cite{Dudarev} , as implemented in VASP. A previous study
 \cite{Faraggi2015}  has already corroborated that the results concerning magnetic moments and 3d level occupations do not change 
appreciably in the 3-5 eV range of the U parameter.

For the geometrical optimization of the free-standing Mn-(F4)TCNQ and Ni-(F4)TCNQ systems, periodic supercell boundary conditions have been imposed. 
The optimal cell dimensions and atomic positions have been obtained by an energy minimization procedure with a convergence 
criterion of  $10^{-6}$ eV for the energy and 0.02 eV/\AA\ for the forces to assure that we reach sufficient accuracy in numerical 
values of the calculated magnitudes. 
The Kohn-Sham wave functions have been expanded in a plane wave basis set with a kinetic energy cutoff of
400 eV for all the systems considered. A Monkhorst-Pack k-point sampling \cite{bib:monk76} equivalent to $8\times12$ in the $1\times1$ surface unit cell and Methfessel-Paxton integration with smearing width 0.1\,eV \cite{bib:methfessel89} have been used. 
Symmetry considerations have been switched off from the calculations
and a preconverged charge density with a fixed value of the total spin for the unit cell has been used to relax all the networks. 
For the obtained relaxed $1\times1$ geometries, where the layer is constrained to be flat, we have evaluated 
the magnetic anisotropy energies with adjusted parameters. 
Total energies have been converged with a tolerance of $10^{-7}$ eV. 
A $12\times18$ k-point sampling and the corrected tetrahedron method of 
integration \cite{bib:bloechl94} have been used instead of smearing methods.

\begin{figure}[hbt]
\centering
\includegraphics[width=0.75\textwidth]{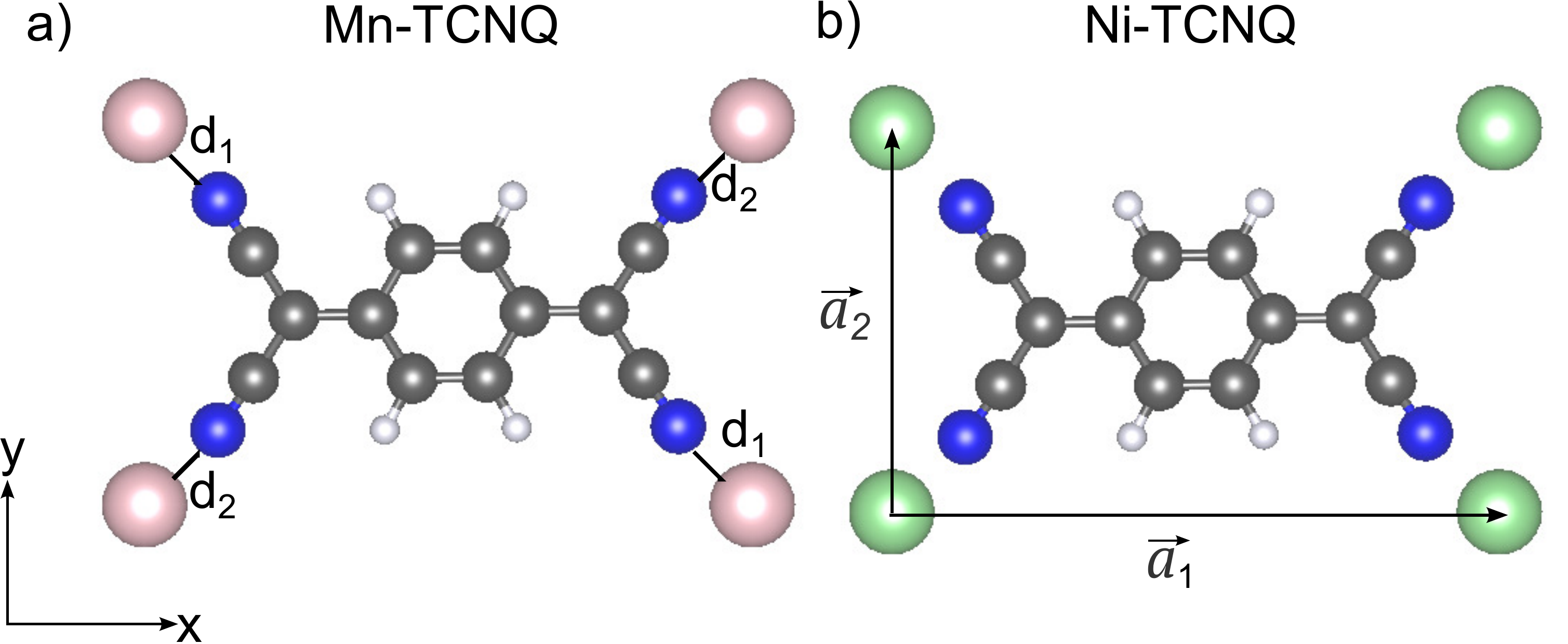}
\caption{ Visualization of the Mn-TCNQ and Ni-TCNQ rectangular cells. Blue, grey and white circles correspond to N, C and H atoms, respectively, while bright violet and bright green circles to Mn and Ni atoms.}
\label{geo}
\end{figure}

Fig.~\ref{geo} shows a top view visualization of the systems considered. The optimized geometrical parameters are included in 
Table~\ref{1x1conf}, where $\vec{a_1} = a_x \hat u_x$ and $\vec{a_2} = a_y \hat u_y$ denote the lattice vectors 
while d$_1$ and d$_2$ denote the values of the Mn-N or Ni-N bond lengths indicated in Figure~\ref{geo}. 

\begin{table}[h]\centering
\begin{tabular}{ccc}
\toprule
\textbf{$1\times1$ cell} &  \textbf{Mn-TCNQ} & \textbf{Ni-TCNQ} \\
\midrule
$a_x$ & 11.52 & 11.32  \\
$a_y$ & 7.38 & 7.16  \\
d$_1$ & 2.12 & 2.01  \\
d$_2$ & 2.12 & 1.95  \\
\bottomrule
\end{tabular}
\caption{ Lattice vectors and bond lengths  (in \AA) 
of the Mn-TCNQ and Ni-TCNQ rectangular cells indicated in Figure~\ref{geo}.}
\label{1x1conf}
\end{table}

\section{Results}

\subsection{XMCD data}
\label{XMCD}

The XMCD intensity variation as a function of the applied magnetic field ($B$) defines a curve that is proportional to the system magnetization. Therefore, when the value of the spin magnetic moments at the metal centers  ($S$) , the temperature ($T$) and the 
Land\'e $g$-factor are known, one can use simple models to simulate the magnetization response. A good reference to be considered is the case of paramagnetic behavior (spins responding individually to the applied magnetic field) that can be represented by a Brillouin function. Whenever a preference for ferromagnetic (FM) or antiferromagnetic (AFM) coupling between spins appears, the corresponding magnetization curves will show higher or lower curvature, respectively, than the corresponding Brillouin function for the same $S$, $T$ and $g$-factor values. In this way, in principle, one can decide about the type of magnetic coupling between localized spins at the metal centers, as long as the value of the spin ($S$) is known. 
Note that, in the presence of strong magnetic anisotropies and high orbital angular moments, the analysis becomes more involved \cite{Gambardella2009}. 
However, here, we can follow this simplified scheme, as shown below. 
According to our DFT calculations, described in section ~\ref{DFT}, Mn atoms in Mn-TCNQ have a localized spin magnetic moment close to $S=5/2$, although somewhat lower, while Ni atoms in Ni-TCNQ have a much lower spin  close to $S=1/2$, although somewhat higher. Therefore, we use the values $S=5/2$ and $S=1/2$ for Mn and Ni, respectively, to do our XMCD analysis that includes fitting curves to XMCD data based on Weiss mean-field theory described in the next section, where $J$ and $D$ are defined, and also a comparison with the corresponding Brillouin functions. 

\begin{figure}[!ht]  
\vspace{0pt}\includegraphics[width=0.45\columnwidth]{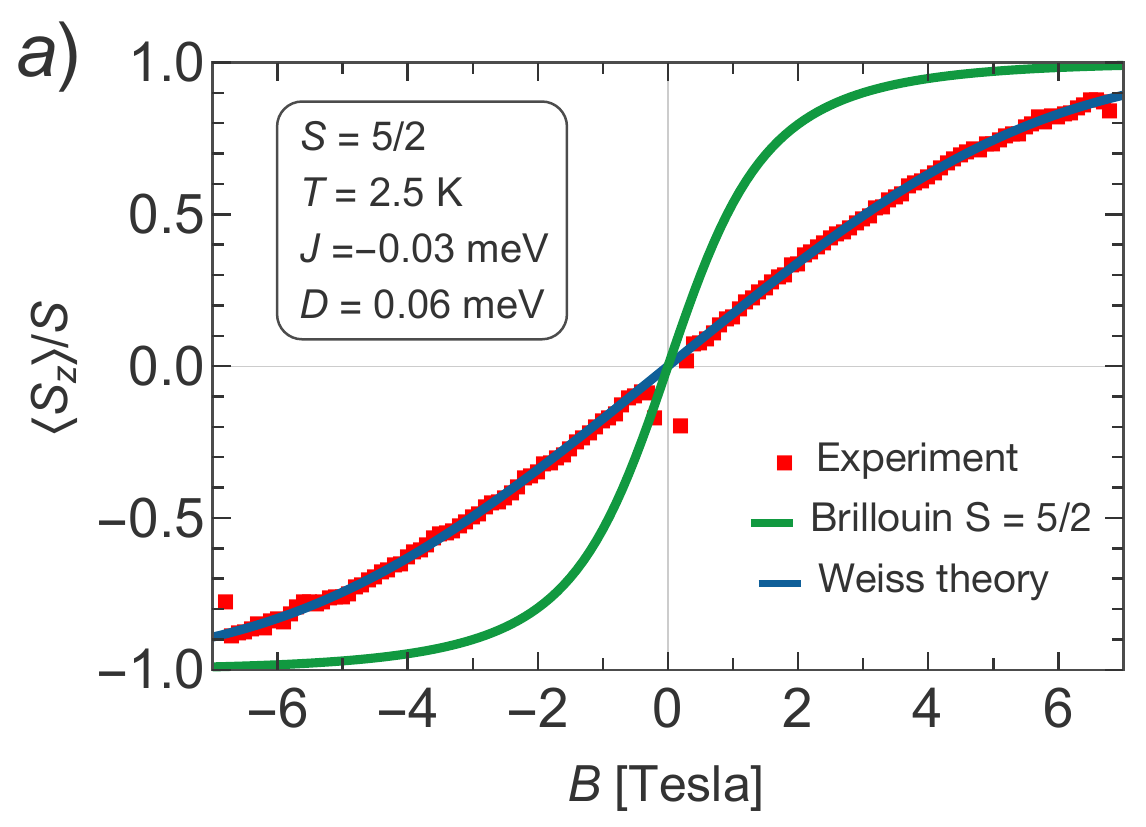}  
\hfill  
\vspace{0pt}\includegraphics[width=0.45\columnwidth]{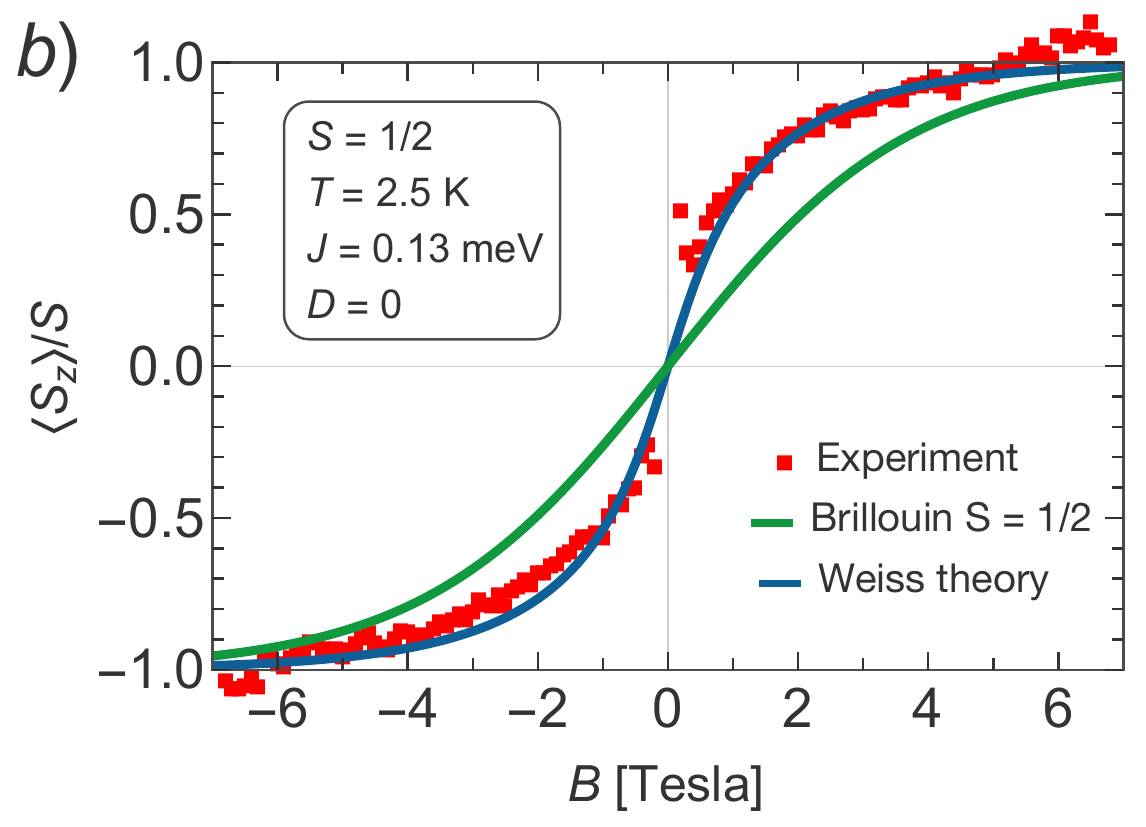}  
\caption{
\label{figureMFfits}  
The best fit with the Weiss mean-field theory to the experimental data
for (a) Mn-TCNQ and (b) Ni-TCNQ at normal beam incidence ($\theta=0\degree$) and 
the temperature $T=2.5\,\textrm{K}$.
The experimental data are shown in red filled squares, whereas the solution of the mean-field self-consistency equations
is shown as the blue solid curve.
For comparison, we also plot the Brillouin function for $S=5/2$ in (a) and $S=1/2$ in (b),
showing that the shape of the measured magnetization versus $B$ deviates substantially from 
the Brillouin function at this temperature.
} 
\end{figure}

\begin{figure}[!ht]  
\vspace{0pt}\includegraphics[width=0.45\columnwidth]{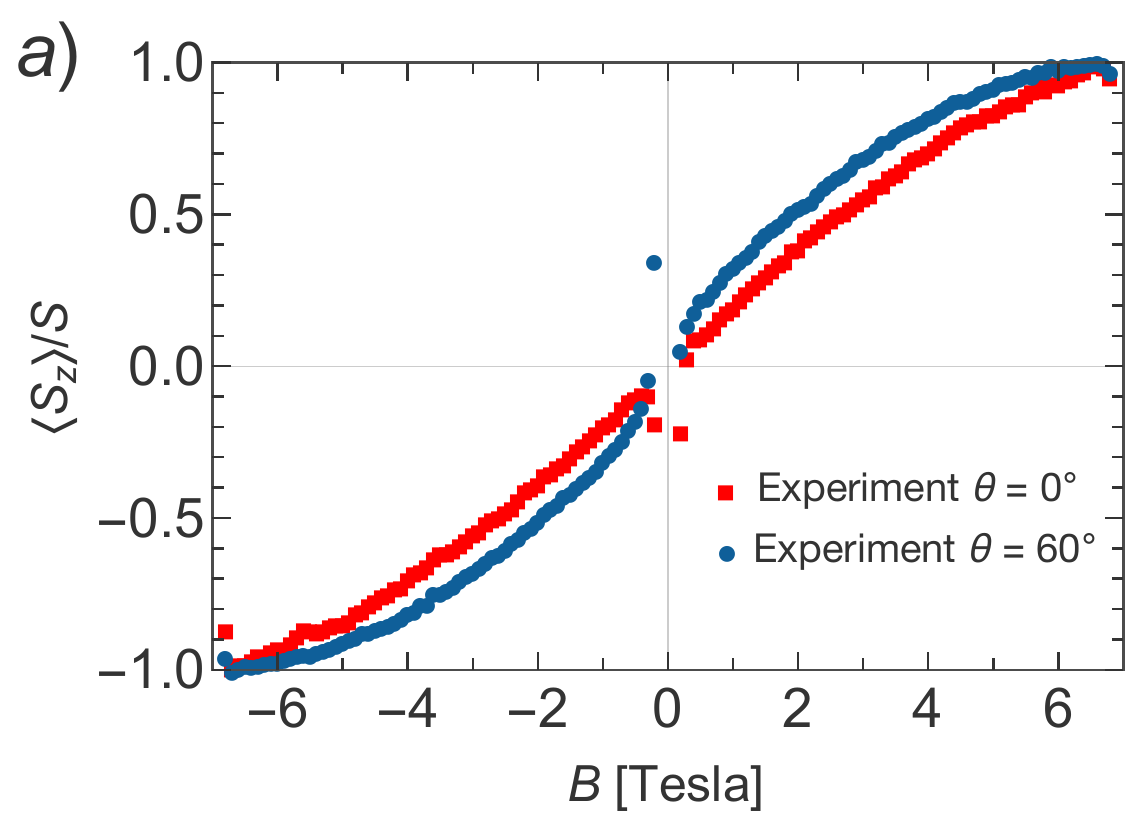}  
\hfill  
\vspace{0pt}\includegraphics[width=0.45\columnwidth]{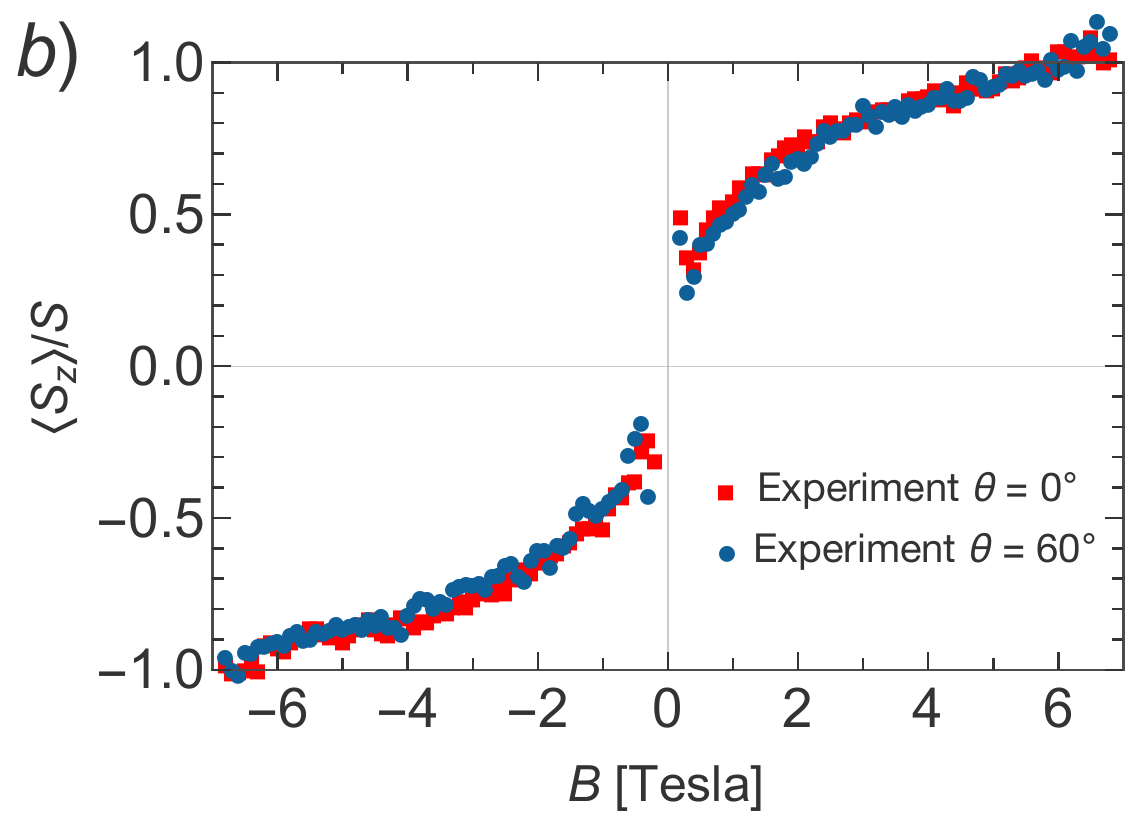}  
\caption{
\label{figureAnisotropy}  
Comparison of the rescaled XMCD signal measured for
(a) Mn-TCNQ and (b) Ni-TCNQ at normal ($\theta=0\degree$) and 
grazing ($\theta=60\degree$) beam incidences.
The data in (a) show a sizable $\theta$-dependence, which we attribute to the single-ion anisotropy for Mn-TCNQ.
In contrast, the data in (b) show no $\theta$-dependence, meaning that there exists no sizable magnetic anisotropy.
}  
\end{figure}

The results are shown in Fig. ~\ref{figureMFfits} a) and b) for Mn-TCNQ and Ni-TCNQ, respectively. It is evident that in Mn-TCNQ the coupling between Mn spins is AFM, while in Ni-TCNQ it is FM. Additionally, the fitted values of the exchange coupling constants reveal a weaker coupling between Mn spins ($J=-0.03 $meV) , as compared to the coupling between Ni spins ($J=0.13$ meV), while the single ion anisotropy parameter $D=0.06$ meV corresponds to a weak anisotropy with in-plane magnetization for Mn-TNCQ and $D=0$ to the absence of anisotropy for Ni-TCNQ. In order to learn more about the magnetic anisotropy of these systems, in Fig.~\ref{figureAnisotropy} we plot a comparison of XMCD data obtained for perpendicular and grazing incidence for Mn-TCNQ and Ni-TCNQ, the former showing a mild angular dependence with stronger intensity for grazing incidence, i. e., a fingerprint of magnetic anisotropy in the system with in-plane magnetization. 
Incidentally, this weak anisotropy is only observed at low temperatures. 
However, in the Ni-TCNQ XMCD data there is no significant angular dependence, which means a negligible magnetic anisotropy. A value of the Ni atom spin $S=1/2$ corresponds to the absence of single ion anisotropy \cite{Dai2008}. 

\subsection{Model for Mn-TCNQ and Ni-TCNQ}
\label{modelH}

\begin{figure}[!ht]
\vspace{0pt}\includegraphics[width=0.45\columnwidth]{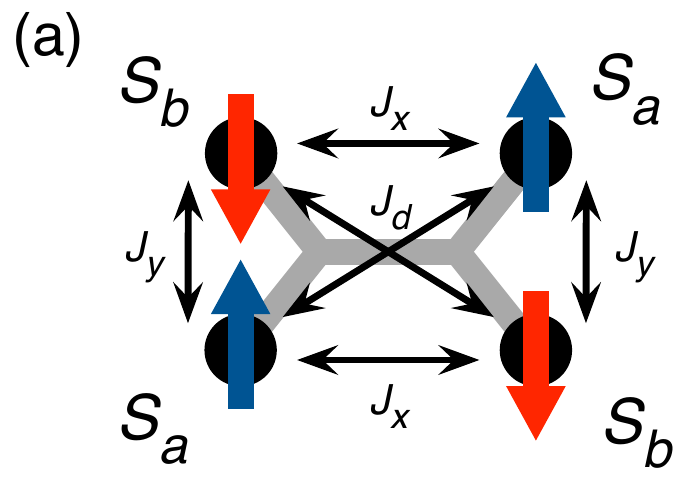}
\hfill
\vspace{0pt}\includegraphics[width=0.45\columnwidth]{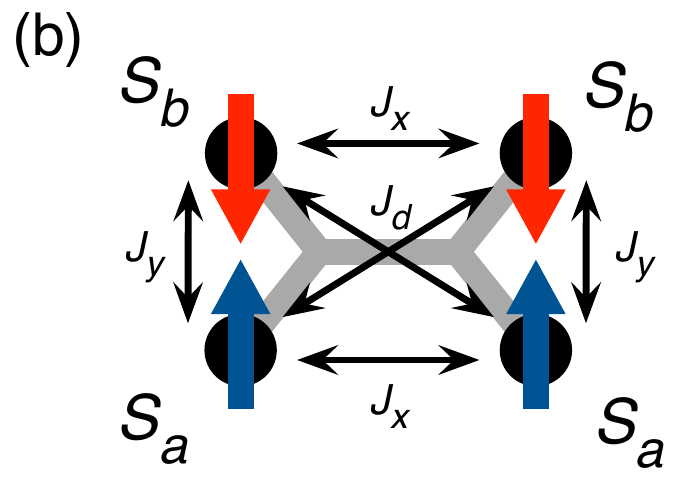}
\caption{\label{figureMnTCNQsuperexchange}
Sketch of the Mn-TCNQ lattice showing the relevant magnetic couplings between the Mn atoms.
The 4-leg TCNQ molecules mediate by superexchange an antiferromagnetic interaction
both between the nearest neighbors on the lattice (couplings $J_x$ and $J_y$) as well as
between the next-to-nearest neighbors (coupling $J_d$). 
For a sufficiently small in magnitude $J_d$, the tendency is to order the spins in the checkerboard pattern (a).
With increasing the magnitude of $J_d$, a crossover to ordering spins in rows or columns takes place (b).
}
\end{figure}

In Mn-TCNQ, the coupling between local moments is antiferromagnetic and occurs by means of the Anderson superexchange mechanism~\cite{PhysRev.179.560, Yosida1996}.
In perturbation theory, the superexchange interaction was found~\cite{Faraggi2015} to be dominated by a virtual process
in which two electrons hop from the doubly occupied LUMO of the TCNQ molecule onto 
two adjacent Mn atoms.
Inclusion of additional molecular orbitals, such as the HOMO, leads to a generic superexchange interaction 
with coupling constants $J_x$, $J_y$, and $J_d$, as shown in Fig.~\ref{figureMnTCNQsuperexchange}.
The model Hamiltonian describing the magnetic properties of Mn-TCNQ, thus, reads
\begin{equation}
H=-\frac{1}{2}\sum_{ij}J_{ij}\bm{S}_i\cdot\bm{S}_j +D\sum_{i}S_{i,z}^2+\textsl{g}\mu_B\sum_{i}\bm{S}_i\cdot\bm{B},
\label{eqHamMnTCNQmodel}
\end{equation}
where $\bm{S}_i$ denotes the local moment of the Mn atom ($S=5/2$) on site $i$, 
$D$ is the single-ion anisotropy energy, $\textsl{g}$ is the 
Land\'e $\textsl{g}$-factor ($\textsl{g}\approx 2$), and $\bm{B}$ is the magnetic field.
The Heisenberg exchange constant $J_{ij}$ is restricted to the nearest ($J_x$ and $J_y$) and next-to-nearest ($J_d$) neighbors on the rectangular lattice.
The summation in the Heisenberg interaction term accounts twice for each pair of interacting sites; 
hence the presence of a factor $1/2$ in Eq.~(\ref{eqHamMnTCNQmodel}).

A quick insight into the tendency to order the spins in this model 
is granted by the Fourier transform of the exchange coupling $J_{ij}$,
\begin{eqnarray}
J_{\bm{q}}&=&\sum_{j}J_{ij}e^{-i\bm{q}\cdot\left(\bm{r}_j-\bm{r}_i\right)}=2 J_x\cos\left(q_xa_x\right)\nonumber\\
&&+2 J_y\cos\left(q_ya_y\right) + 
4J_d\cos\left(q_xa_x\right)\cos\left(q_ya_y\right),
\end{eqnarray}
where $\bm{q}=(q_x,q_y)$ is the two-dimensional wave vector and 
$\bm{r}_i$ is the position of the Mn atom on site $i$.
For ferromagnetic couplings ($J_{ij}>0$), the maximum of $J_{\bm{q}}$ occurs at $q=0$,
which indicates that the spin order could be uniform from a mean-field point of view, 
not addressing the question about its stability
against fluctuations in two dimensions. 
Additional terms, such as the single-ion anisotropy or the Zeeman interaction,
may stabilize the uniform spin order.

In contrast, for antiferromagnetic couplings ($J_{ij}<0$), the maximum of $J_{\bm{q}}$ occurs usually at the edge
of the Brillouin zone, indicating that the magnetization is staggered in some way over the unit cell. 
When only nearest neighbors are coupled ($J_x=J_y\neq 0$ and $J_d=0$), the maxima lie at 
$\bm{q}=(\pi/a_x,\pi/a_y)$ and its equivalent points, which results in the usual 
checkerboard-like antiferromagnetic order [see Fig.~\ref{figureMnTCNQsuperexchange}\,(a)]. 
As the diagonal coupling is turned on (assuming an antiferromagnetic $J_d<0$), 
for a sufficiently large magnitude of $J_d$ there is a transition from 
the checkerboard pattern to a so-called \emph{superantiferromagnetic} state 
of antiferromagnetically ordered rows or columns. 
For $\left|J_y\right|>\left|J_x\right|$, by 
requiring $\partial J_{\bm{q}}/\partial q_x\equiv 0$ at $q_y=\pi/a_y$,
we find at $J_d=J_x/2$ the transition point for antiferromagnetic column formation 
[see Fig.~\ref{figureMnTCNQsuperexchange}\,(b)].

The effect of the diagonal coupling $J_d$ consists in introducing magnetic frustration~\cite{PhysRev.179.560,PhysRevB.39.2344} in the spin lattice.
We remark here that the special point $J_d=J_x/2$ is realized to a good approximation in our Mn-TCNQ lattice, because 
(i) the LUMO of the TCNQ molecule has 
overlap with the $d_{zx}$ and $d_{yz}$ orbitals of the Mn atom, thus,
dominating the superexchange,
and (ii) the direct coupling between the LUMOs of neighboring TCNQ molecules is rather weak.
The latter makes it possible to consider two independent paths of superexchange for the nearest neighbors, 
with each path going separately via one of the two TCNQ molecules connecting the two neighboring Mn atoms.
For the diagonal coupling, only one path is possible, which leads to a reduction of the diagonal coupling by a factor of $2$ 
as compared to the nearest-neighbor coupling. With the approximations (i) and (ii), the coupling constants obey 
$J_x=J_y=2J_d$ (see Ref.~\cite{Faraggi2015} for further details).

Despite the fact that the Mn-TCNQ lattice 
may well be in a frustrated magnetic state consisting of a mixture of the two phases in Fig.~\ref{figureMnTCNQsuperexchange},
the XMCD data appear to be consistent with a much simpler description of the magnetization as a function of the $B$-field,
which is derived from the Weiss mean-field theory, and it captures faithfully weak deviations from the paramagnetic state.
The superexchange couplings are rather weak,~\cite{Faraggi2015} of the order of $10^{-5}\,\textrm{eV}$, and the Zeeman term soon dominates.   
Additionally, there exists a fair amount of single-ion anisotropy, described by the $DS_z^2$ term in Eq.~(\ref{eqHamMnTCNQmodel}).

We make the mean-field approximation for the model in Eq.~(\ref{eqHamMnTCNQmodel}),
\begin{eqnarray}
&&H\approx H_{\textrm{MF}} := H_{\textrm{loc}}
+\frac{1}{2}\sum_{ij}J_{ij}\left\langle\bm{S}_i\right\rangle\cdot\left\langle\bm{S}_j\right\rangle,\nonumber\\
&&H_{\textrm{loc}}=\sum_i\bm{S}_i\cdot \bm{h}_i+D\sum_{i}S_{i,z}^2,\nonumber\\
&&\bm{h}_i=\textsl{g}\mu_\textrm{B}\bm{B}
-\sum_{j}J_{ij}\left\langle\bm{S}_j\right\rangle,
\label{eqHlochi123}
\end{eqnarray}
where $H_{\textrm{loc}}$ gives the local description of the interacting system in terms of the Weiss fields $\bm{h}_i$.
The spin averages $\left\langle\bm{S}_i\right\rangle$ can be regarded as variational parameters of the theory.
The last term in the first line of Eq.~(\ref{eqHlochi123}) compensates for the double counting of interaction energy occurring 
in the local Hamiltonian $H_{\textrm{loc}}$ and plays an important role when calculating the free energy of the interacting system.
The minimization of the free energy allows us to determine the values of the order parameters  $\left\langle\bm{S}_i\right\rangle$. 
The procedure is described in the Appendix.

Next, we focus on the  XMCD data taken at normal incidence ($\theta=0\degree$), 
for which the magnetic field is applied along the $OZ$-axis,
$\bm{B}=(0,0,B)$. For the (checkerboard) antiferromagnetic phase,
we use two order parameters $S_a$ and $S_b$, which represent the $Z$-components 
of the spins in the unit cell as shown in Fig.~\ref{figureMnTCNQsuperexchange}\,(a), 
and minimize the upper bound to the free energy [$F_{\textrm{AF}}(S_a,S_b)$] with respect to the order parameters $S_a$ and $S_b$.
Alternatively, one can require stationarity of free energy, $\partial F_{\textrm{AF}}/\partial S_a=0$ 
and $\partial F_{\textrm{AF}}/\partial S_b=0$,
which yields two coupled equations,
\begin{equation}
S_a=\frac{\partial F_1(h_a)}{\partial h_a}\quad\quad\mbox{and}\quad\quad S_b=\frac{\partial F_1(h_b)}{\partial h_b},
\label{eqSaSbselfconsistAF}
\end{equation}
where $F_1$ is the free energy of a single isolated spin.
The mean-field solution is obtained from these self-consistent equations.
As a rule, several solutions are found.
The choice of the physical solution relies again on the least value of the free energy. 
For the superantiferromagnetic phase, we use again two order parameters, $S_a$ and $S_b$, but now they are distributed in the unit cell 
as shown in Fig.~\ref{figureMnTCNQsuperexchange}\,(b).
The mean-field approximation takes into account only the connections (\emph{i.e.}\ bonds) 
between the spins on a local scale, whereas 
the constrains related to the dimensionality of the systems go unaccounted.
We can, therefore, adapt here all the results derived for the phase in Fig.~\ref{figureMnTCNQsuperexchange}\,(a)
by simultaneously replacing $J_x$ and $J_d$ in all expressions as 
\begin{equation}
\left\{\begin{array}{c}
J_x\to 2J_d,\\
J_d\to J_x/2.
\end{array}\right.
\label{eqsubstJxJd}
\end{equation}
The factors $2$ and $1/2$ appear here because 
each $J_x$ connector counts as half a bond in the unit cell, whereas 
each $J_d$ connector counts as a full bond.

We fit the experimental data for normal magnetic fields in Fig.~\ref{figureMFfits} 
assuming the relation $J_x=J_y=2J_d$, which corresponds to the case when a single orbital of the ligand 
is dominating the superexchange. We reach a good fit to the experimental data
for $J_x=-0.02\,\textrm{meV}$. Our working assumption was that the critical temperature ($T_N^{\textrm{Weiss}}$) 
is sufficiently low as to allow application of the Weiss theory, \emph{i.e.}\ $T_N^{\textrm{Weiss}}<T$.
This means also that the order parameters $S_a$ and $S_b$ are never of opposite sign 
and are, in fact, equal to each other over the full range of applied magnetic fields.
Therefore, the experimental data can equally well be fitted by a ferromagnetic mean-field theory with antiferromagnetic coupling constants.
To simplify the matter even further, we consider a square lattice with a single coupling constant $J$.
Effectively, this coupling constant will be related to the previous coupling constants by equating to each other $J_{\bm{q}}$ at $q=0$ 
for both models, which immediately yields $4J=2J_x+2J_y+4J_d$. 
Using the above value, we arrive at $J=3J_x/2=-0.03\,\textrm{meV}$ and  $D=0.06\,\textrm{meV}$ for $S=5/2$.

The same effective model derived from a mean field Hamiltonian with $J$, $S$ and $D$ parameters  can be used for Ni-TCNQ, although its relation with the microscopic Hamiltonian described in Ref. \cite{Faraggi2015}  is different. In this case, we find a good fit with  $J=0.13\,\textrm{meV}$ and $D=0$ for $S=1/2$.

\subsection{Spin polarized DFT+U calculations}
\label{DFT}

\begin{figure*}
 \includegraphics
[width=0.9\columnwidth]  
 {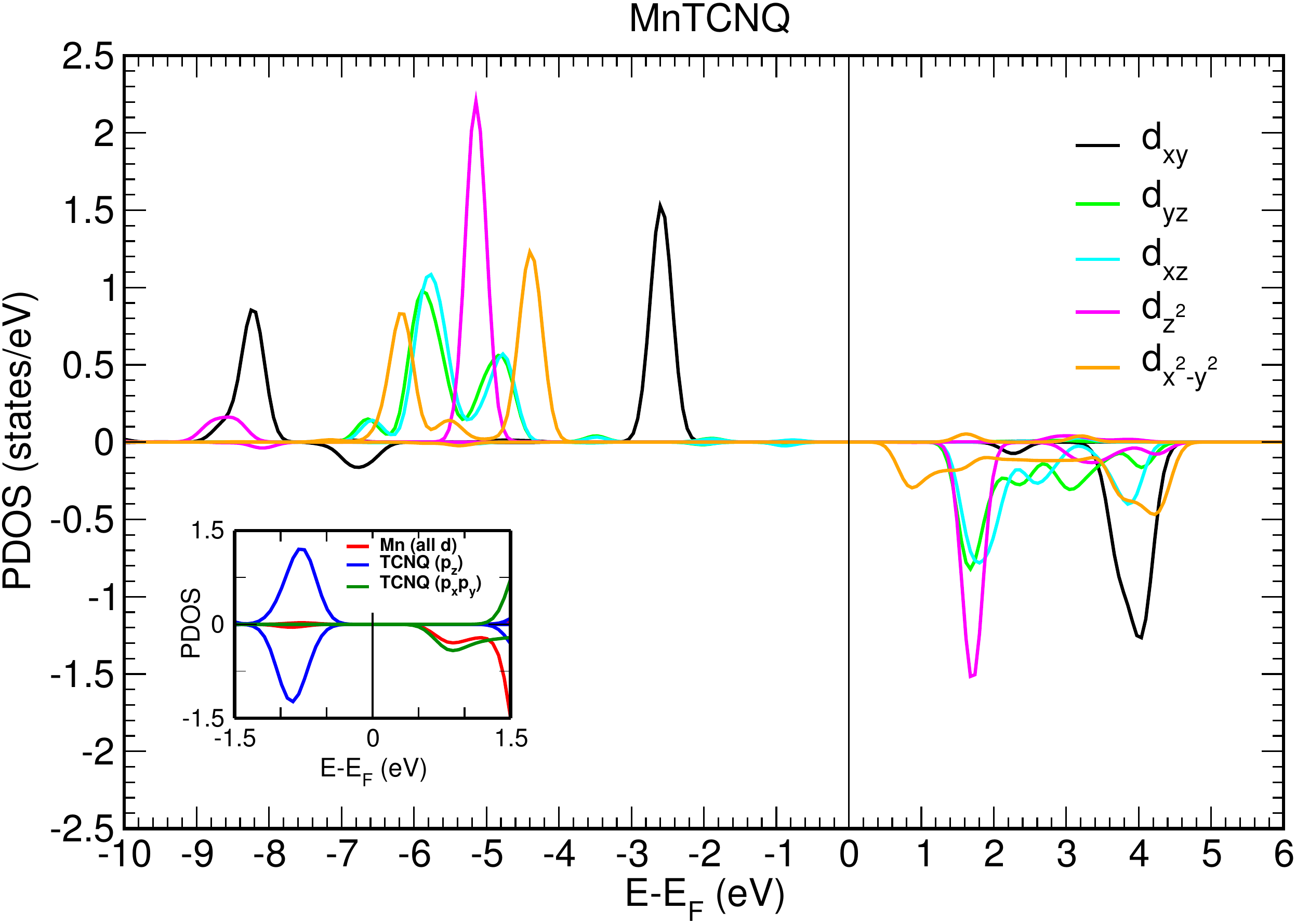}
\caption{{Projected density of states (PDOS) onto the five different Mn($3d$) orbitals for Mn-TCNQ. The inset shows the PDOS onto $p$ orbitals of C and N atoms in TCNQ, as well as onto all Mn($3d$) orbitals, in a narrow energy range close to the Fermi level ($E_F$).
}}
 \label{PDOS_Mn_d}
\end{figure*}

\begin{figure*}
 \includegraphics
[width=0.9\columnwidth]  
 {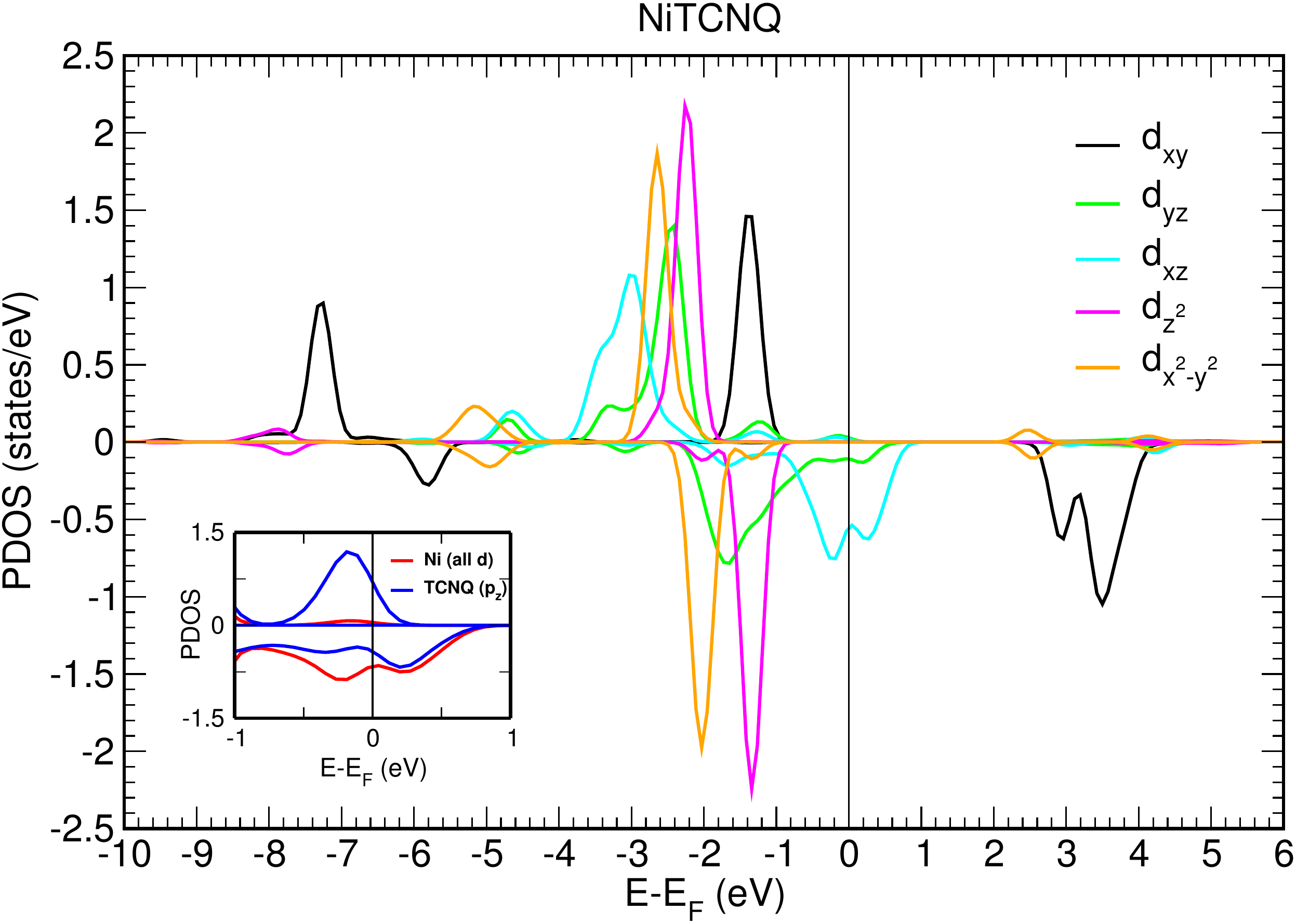}
\caption{{Projected density of states (PDOS) onto the five different Ni($3d$) orbitals for Ni-TCNQ. The inset shows the PDOS onto $p$ orbitals of C and N atoms in TCNQ, as well as onto all Ni($3d$) orbitals, in a narrow energy range close to the Fermi level ($E_F$).
}}
 \label{PDOS_Ni_d}
\end{figure*}

We first consider a two-dimensional free-standing overlayer description for Mn-TCNQ and Ni-TCNQ networks. Both the lattice vectors and atomic positions have been optimized by using an energy minimization procedure within DFT, as described in the Materials and Methods section. The projected densities of states (PDOS) onto different atomic $3d$ orbitals of the Mn and Ni atoms are shown in Figs.~\ref{PDOS_Mn_d} and ~\ref{PDOS_Ni_d}, respectively. The insets show the PDOS onto atomic $p$ orbitals of the C and N atoms of the organic ligand, as well as onto Mn and Ni 3d states without m number resolution, in a narrow energy range close to the Fermi level. A close inspection of Figs.~\ref{PDOS_Mn_d} and~\ref{PDOS_Ni_d}  reveals important differences between the two systems under study. The most significant is the half-filling of the $3d$ states with all the majority spin states occupied in Mn-TNCQ, which corresponds to a value of the spin approximately equal to $S=5/2$ localized at the Mn atoms, while in Ni-TCNQ only one minority spin state is fully unoccupied ($3d_{xy}$),  
which corresponds to a value of the spin localized at the Ni atom approximately $S=1/2$, although somewhat higher, as the minority spin states $3d_{xz}$ and $3d_{yz}$ are partially occupied. Additionally, in Ni-TCNQ, the $3d_{xz}$ and $3d_{yz}$ states are hybridized with TCNQ orbitals close to the Fermi level, in particular the LUMO, giving rise to a delocalized spin density \cite{Faraggi2015}. This can be seen by comparing the PDOS onto atomic $p$ orbitals of the C and N atoms of the TCNQ organic ligand shown in the insets of Figs.~\ref{PDOS_Mn_d} and ~\ref{PDOS_Ni_d} for Mn-TCNQ and Ni-TCNQ, respectively. In Ni-TCNQ the LUMO orbital is spin polarized but this is not the case in Mn-TCNQ, for which the TCNQ LUMO practically does not hybridize with Mn states and it is fully occupied.

Next,  using these two optimized structures calculated with a $1\times1$ surface unit cell within the DFT+U method with spin polarization as starting point, we proceed to double the size of the surface unit cell into a $2\times1$ cell that contains two metal centers (Mn or Ni atoms)  and two TCNQ molecules. In this way, we can decide which is the most favorable type of magnetic coupling ( ferro- or antiferro-magnetic) between spins localized at the Mn or Ni centers by comparing the values of the corresponding total energies. We consider a checkerboard configuration using oblique vectors in the $2\times1$ surface unit cell and confirm that ferromagnetic coupling is favorable in Ni-TCNQ networks, while in Mn-TCNQ networks antiferromagnetic coupling is preferred in agreement with Ref. \cite{Faraggi2015}. The corresponding spin densities are shown in Fig.~\ref{spin_density} for Mn-TCNQ and Ni-TCNQ. 
In section II of the Supplementary Material we also include other configurations obtained by using a rectangular $2\times2$ surface unit cell, in which other AFM configurations with spins aligned in rows or columns are considered as well  \cite{Otrokov2015}, showing the importance of next to nearest neighbors (diagonal) couplings in the networks that has been discussed in the previous section.

\begin{figure*}
 \includegraphics
 [width=\textwidth]
 {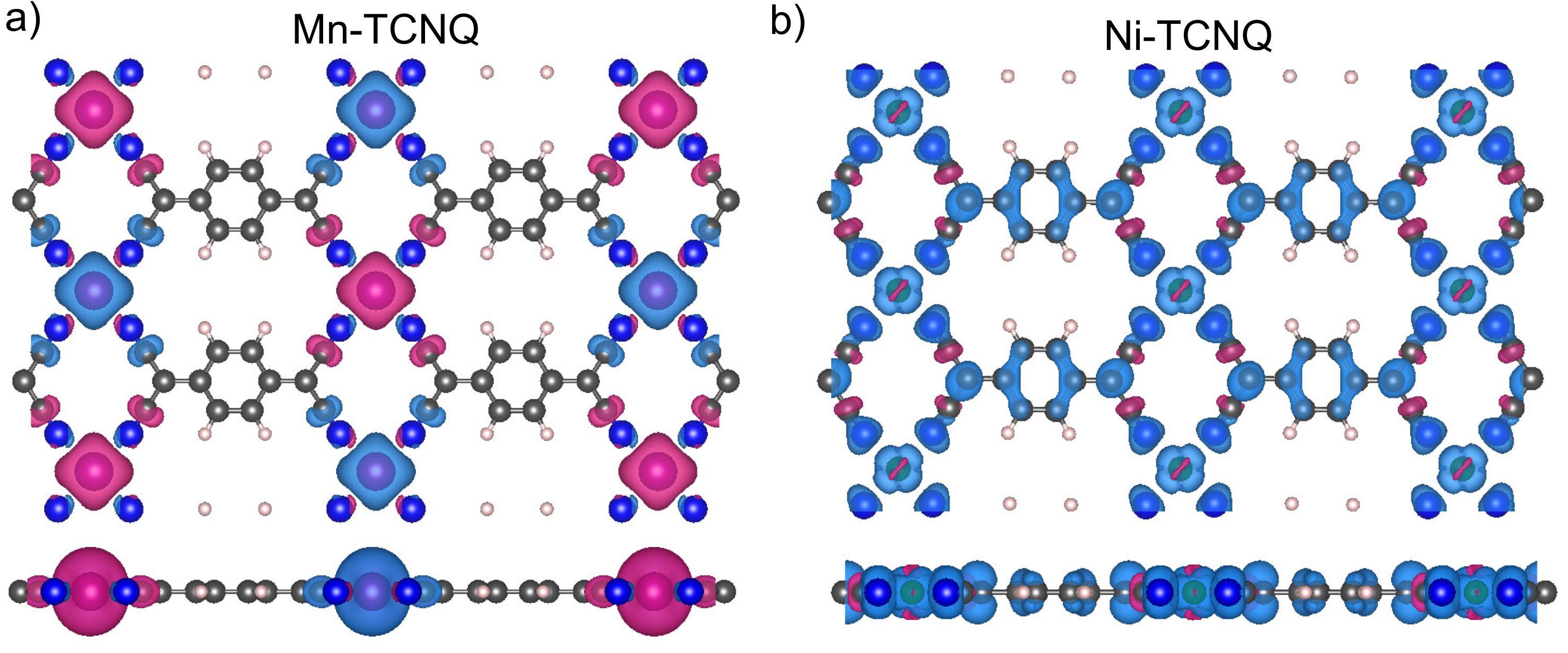}
\caption{{Top (upper panels) and side (lower panels) views of the calculated spin-densities for a) Mn-TCNQ and b) Ni-TCNQ free standing overlayers.
}}
 \label{spin_density}
\end{figure*}

\subsection{Magnetocrystalline anisotropy}
\label{MAE}

The magnetocrystalline anisotropy energies (MAE) can be obtained from DFT calculations
that include SOC effects. 
The resulting total energies, thus, depend on the orientation of the magnetization density.
For extended systems, where the transition metal atomic orbital momentum is expected to be 
partially or totally quenched, the MAE appears as a second order SOC effect. 
In systems where the PDOS is characterized by sharp peaks and devoid of degeneracies
at the Fermi level, a second-order perturbative treatment of the SOC 
makes it possible to establish a few guidelines for the likelihood of an
easy axis or plane. 
The perturbation couples states above and below the Fermi
level and it is inversely proportional to the energy difference between states.
When the spin-up $d$-band is completely filled, it can be shown that the energy
correction is proportional to the expected value of the orbital magnetic moment
and that the spin-flip excitations are negligible
\cite{bib:bruno89,bib:vdlaan98,bib:stoehr99}.

The total energy variation as a function of the magnetization 
axis direction is very subtle, often in the sub-meV range per atom.
When spin-orbit effects are not strong, it is common practice to use the 
so-called second variational method \cite{bib:li90}, where SOC is not treated 
self-consistently. First, a charge density is converged in a collinear 
spin-polarized calculation. Next, a new Hamiltonian that includes a SOC term 
is constructed and diagonalized for two different 
magnetization directions. Then, the MAE is calculated from the difference 
of the two band energies. Alternatively, a more precise MAE can be obtained from 
total energy calculations that include SOC self-consistently. 
Using the latter method, in this work we have calculated 
MAE values for free-standing Mn-TCNQ and Ni-TCNQ networks.

The small energies involved in the anisotropy are a challenge for DFT calculations.
The MAE is highly sensitive to the geometry and electronic structure calculation details,  
such as the exchange and correlation functionals and basis set types. 
From a technical perspective, a reliable MAE is only achieved with demanding 
convergence criteria. For example, it has been observed 
that fine k-point samplings of the Brillouin zone are needed \cite{bib:abdel09,bib:khan16,Otrokov.2dm2017}. 
An account of the convergence details as well as MAE dependence on the $U$ parameter
can be found in section III of the Supplementary Material.

Table~\ref{tab:mae} shows the obtained values for $U=4$\,eV in $1\times1$ cells 
(i.e. , only ferromagnetic ordering is considered in this Section).
For the Mn-TCNQ rectangular network, we find in-plane magnetization with negligible 
azimuthal dependence, i.e., easy plane anisotropy. 
The MAE, calculated as the total energy difference 
between magnetic configurations with Mn magnetizations parallel to the 
OX and OZ axes, is 0.2\,meV.
In the Ni-TCNQ rectangular network the energetically preferred magnetization is out-of-plane 
and the MAE values vary significantly with the azimuthal direction.
As shown in Table~\ref{tab:mae}, the values change as much as 1.50\,meV with azimuthal angle variations. 

\begin{table}[H]
\caption{
Magnetocrystalline anisotropy energies (in meV) for Mn- and Ni-TCNQ calculated as
the difference $\mathrm{MAE}=E_{tot}(0,0)-E_{tot}(90^\circ,\phi)$,
where the two values in parenthesis are the polar and azimuthal angles,
respectively, defining the magnetization direction. Positive (negative)
energies indicate in-plane (out-of-plane) anisotropy.
The last line corresponds to the oblique cell Ni-TCNQ model with angle $\gamma=77.43^\circ$, 
where the anisotropies at the directions of the long (short) pair of Ni-N
bond directions are shown.
The table values have been obtained for $U=4$\,eV with an
energy cutoff of 400\,eV and a $12\times 18\times 1$
k-point sampling, using the tetrahedron method for integration.
}
\label{tab:mae}
\centering
\begin{tabular}{cccccc}
\toprule
\textbf{}     & \textbf{$\phi=0$}   & \textbf{$\phi=90^\circ$} & \textbf{$\phi=45^\circ$} & \textbf{$\phi=-45^\circ$} \\
\midrule
Mn            &  0.20 &  0.19 &  0.20 & 0.20 \\
Ni            & -1.44 & -0.95 & -1.95 & -0.45 \\
\midrule
\textbf{}     & \textbf{$\phi=0$} & \textbf{$\phi=90^\circ$} & \textbf{$\phi=22.6^\circ$} & \textbf{$\phi=-57.4^\circ$} \\
Ni (oblique)  & -0.07 & -0.04 & 0.03 & -0.09 \\
\bottomrule
\end{tabular}
\end{table}

The different behavior of the MAE with the azimuthal angle in Mn and Ni networks can be 
understood in terms of the differences in the metal-molecule bonds, particularly the Mn-N and Ni-N bonds. 
In both networks the $d_{x^2-y^2}$ (with magnetic quantum number $m=2$), $d_{xy} (m=-2)$, and $d_{z^2} (m=0)$ 
orbitals remain rather localized, whereas the $d_{xz} (m=1)$ and $d_{yz} (m=-1)$ orbitals 
are spread over a wider energy range of a few eV below the Fermi level 
(see Figs.~\ref{PDOS_Mn_d} and ~\ref{PDOS_Ni_d}).
The delocalization of electronic charge in these $d_{xz}$ and $d_{yz}$ orbitals 
is stronger in the Ni-TCNQ case, where the latter two sub-bands 
are partially occupied and form hybrid states at the Fermi level with the TCNQ LUMO. 
As these hybrid states lie at the Fermi level, they have a 
dominant role in the magnetic anisotropy and, since they yield markedly directional 
charge and spin density distributions along the Ni-N bonds, they are likely to produce 
azimuthal MAE variations. Conversely, the Mn $d$-electrons hybridize 
weakly with the TCNQ orbitals close to the Fermi level, i. e., with the LUMO, 
and have essentially no weight at the Fermi level. 
The spatial extent of these relevant Ni-TCNQ hybrid states is manifested in the 
delocalized electron spin densities depicted in Fig.~ \ref{spin_density} b), 
as compared to the case of Mn-TCNQ shown in Fig.~ \ref{spin_density} a) 
with a spin density more localized at the Mn sites and its neighboring cyano groups.

The existence of an easy axis (plane) of magnetization for Ni (Mn) cannot be 
anticipated from the electronic structure details. 
In the Mn-TCNQ system, since the $d$-band is half-filled, 
the MAE is led by spin-flip excitations and, therefore, the value of the 
exchange splitting is determinant. 
In the absence of same-spin excitations, the anisotropy would be associated to
the anisotropic part of the spin distribution. More precisely, the 
MAE would be proportional to the anisotropy of the expected values of 
the magnetic dipole operator \cite{bib:vdlaan98,bib:stoehr99}.
However, Fig.~\ref{spin_density} a) shows an anisotropic spin distribution 
extended towards the cyano groups of the organic ligand TCNQ in the network plane by the crystal field. 
The quadrupolar moment of this distribution
should promote out-of-plane magnetization. This interpretation is at variance with the 
SOC-self-consistent DFT result. 
A more elaborated model has been proposed for systems with localized $d$-orbitals. 
It states that the spin-flip excitations that keep the quantum number
$|m|$ constant favor an in-plane magnetization \cite{bib:ke15}. 
The calculated PDOS of Fig.~ \ref{PDOS_Mn_d} shows that the two $|m|=2$ peaks 
($d_{xy,\uparrow}-d_{x^2-y^2,\downarrow}$) are those closer to the Fermi level, 
for majority and minority spin states, respectively. 
This situation is, in principle, compatible with an easy plane behavior.
The conclusion we draw is that the basic qualitative feature of the 
magnetic anisotropy, namely the magnetization direction, 
cannot be accounted for by rules of general character, not even in a case like Mn-TCNQ, 
where the $d$-electrons have a rather localized character that would make 
this system seem \emph{a priori} a good playground for these models.

Next, we turn our attention back to the case of Ni-TCNQ, where the DFT calculations
yield a relatively large value for the MAE with out-of-plane magnetization, as well as significant
variations of the MAE in the network plane.
This theoretical result contrasts with 
the experimental absence of anisotropy in this system and, thus, requires a further analysis oriented 
at finding an explanation. As we discuss below, 
the discrepancy could be explained by substrate effects, mostly due to electronic charge transfer from the metal Au(111) surface. 
However, if we tried to calculate MAE values from DFT calculations with SOC
using the supported Ni-TCNQ/Au(111) model structures presented above, 
we would not obtain informative results, since it would be very difficult in practice to 
disentangle the anisotropy effects originated by different aspects of the system. The most significant of them is the unavoidable artificial strain 
introduced in the system by forcing a commensurable Ni-TCNQ overlayer on top of the Au(111) surface due to the use of periodic boundary conditions 
in a finite size system imposed by our DFT calculations.
However, these limitations can be more conveniently understood using free-standing models.

In the $1\times1$ rectangular Ni-TCNQ free-standing overlayer  we can attribute the large 
MAE values to the partially occupied Ni($d_{xz},d_{yz}$) states. If these $|m|=1$ 
bands were completely filled by transfer of 0.5\, electrons from the metallic substrate, 
their contribution to the MAE would be dramatically reduced. 
Additionally, the Ni atom spin would become close to $S=1/2$, a case for which no single-ion anisotropy is possible \cite{Dai2008}. 
However, it is hard to give a precise 
estimate of the amount of charge transfer and, on top of this, other sources of anisotropy 
reduction could be at play, like a reduction of Ni coordination due to a geometrical distortion.
Indeed, the lowest-energy configuration of this rectangular unit cell is obtained 
upon a small symmetry-lowering distortion where the four Ni-N bonds are inequivalent:
the bonds at $45^\circ$ degrees with the OX axis (d$_2$) have a length of 1.95\,{\AA} and 
the other pair at $-45^\circ$ (d$_1$) of 2.01\,{\AA}  (see Fig.~\ref{geo}). 
The former direction is that of the hardest magnetization axis. 
This symmetry breaking, though subtle from the geometry point of view, 
is nevertheless associated to a noticeable asymmetry in the electronic structure, 
which is in turn behind the strong azimuthal variability of the MAE.
In a closer inspection  of the PDOS we find that the Ni($d_{xz},d_{yz}$) peaks at the Fermi level 
hybridized with the molecule LUMO are contributed by $d$-orbitals lying on 
the plane containing the short Ni-N bonds (d$_2$) 
and the surface normal (see section III of the Supplementary Material).
The long bonds (d$_1$), to which $|m|=1$ states at the Fermi level do not contribute , 
correspond to a softer magnetization direction.

To understand the consequences of this distorted geometry on the magnetic anisotropy, 
we have constructed a free-standing flat Ni-TCNQ model in an oblique unit cell, 
in which the angle $\gamma$ between the lattice vectors $\vec{a_1}$ and $\vec{a_2}$ 
is varied  (the rectangular cell corresponds to $\gamma = 90^\circ$).
As described in section III of  the Supplementary Material, two cases have been considered: 
 a weakly distorted case with $\gamma = 83.5^\circ$ and a larger distortion with $\gamma = 77.43^\circ$. 
The unit cell angle $\gamma$ has been reduced while uniformly scaling the lattice 
constants to keep the unit cell area equal to that of the rectangular equilibrium unit cell. 
Then, the atomic $(x,y)$ coordinates have been relaxed to satisfy the same 
convergence criteria as in other models of the present work. 
For a larger distortion of the rectangular cell with $\gamma = 77.43^\circ$, 
one could force a commensurate supercell $[(5,2), (1,3)]$ on Au(111) \cite{Faraggi2012}. 
In the optimized structure the TCNQ is barely deformed, 
but one Ni-N bond at the azimuthal direction $\phi=22.6^\circ$ 
is broken because of the cell distortion and the pair of bonds at the $\phi=-75.4^\circ$ 
direction have their lengths reduced to 1.85\,{\AA}  (see section III of  the Supplementary Material). 
The magnetic anisotropy is significantly reduced with respect to that of the 
rectangular cell, but the hardest direction is still the one along the shortest 
pair or Ni-N bonds (see Table~\ref{tab:mae}). 
The main consequence of the Ni coordination reduction caused by the cell shape change is to 
partially quench its spin. We observe that the local magnetic moment 
is reduced by about $0.3 \mu_B$, approaching the ideal $S=1/2$ state that would yield no anisotropy 
in the single-atom picture. 
We observe, nevertheless, that this distorted configuration still has partially 
filled Ni $d_{xz,yz} (|m|=1)$  states at the Fermi level (see section III of  the Supplementary Material).
Thefore, we note that this mechanism of anisotropy reduction and the charge transfer 
effect proposed above are of a different nature, albeit both originate from the interaction 
with the substrate.

All in all, the observed lack of magnetic anisotropy in the Ni-TCNQ/Au(111) XMCD data is clearly a 
substrate effect, which reduces the Ni-TCNQ anisotropy by a combined effect of 
charge transfer and change of coordination. Nonetheless, other subtle 
substrate effects not considered here might also have a role, such as 
fluctuations in the Au-Ni charge transfer due to the incommensurability 
and corrugation of the network.

\section{Discussion and Conclusions}
\label{DandC}

Motivated by the XMCD data, we have performed a thorough analysis of the magnetic properties that characterized Mn and Ni metal-organic coordination networks, focusing on the magnetic coupling and anisotropy. By fitting the XCMD data using a model Hamiltonian based on mean-field Weiss theory and comparing with Brillouin functions, we  find a completely different behavior for Mn and Ni networks: while in Mn networks the spins localized at the Mn centers are coupled antiferromagnetically with a mild preference to in-plane magnetization, in Ni networks the spins localized at the Ni atoms are coupled ferro-magnetically and do not show any sizable magnetic anisotropy. 

These observations are also rationalized with the help of density functional theory calculations in two steps: first we focus on the magnetic coupling and next we address the subtle question of the magnetic anisotropy. Spin-polarized DFT calculations using a $1\times1$ surface unit cell to describe the free-standing-overlayers reveal a very different electronic structure close to the Fermi level for the two systems under study. The system Mn-TCNQ is insulating and has weak hybridization between Mn and TCNQ states close to the Fermi level, while in Ni-TCNQ hybridization between Ni(3d) states and the TCNQ LUMO at the Fermi level is rather significant. This difference permits to explain the observed trends in XMCD data with antiferromagnetic (ferromagnetic) coupling for Mn (Ni) networks that is also confirmed by another set of DFT calculations using a $2\times1$ surface unit cell. 

We find that the basic qualitative feature of the magnetic anisotropy, namely the magnetization direction, cannot be accounted for by rules of general character.  Actually, the magneto-crystalline anisotropy is contributed by many electron excitation channels and it clearly shows an intricate dependence on the fine electronic structure details of each particular system. While in Mn-TCNQ/Au(111) the observed magnetic anisotropy with in-plane magnetization agrees with the DFT calculations for the neutral Mn-TCNQ overlayer, the observed lack of magnetic anisotropy in Ni-TCNQ/Au(111) suggests the existence of a substrate effect, which reduces the Ni-TCNQ anisotropy due to a combination of electronic charge transfer and change of Ni-N coordination.






\vspace{6pt} 

\supplementary{The following are available online at www.mdpi.com/link; I. Supplementary XAS and XMCD data for TCNQ and F4TCNQ networks, II. Energetics of different magnetic configurations using a 2x2 unit cell, III. MAE convergence details, dependence with U and with lattice distortions.}

\acknowledgments{We thank MINECO and the University of the Basque Country (UPV/EHU) for partial financial support, grant numbers FIS2016-75862-P and IT-756-13, respectively, the first one covering costs to publish in open access journals.
MMO acknowledges the support by the Tomsk State University competitiveness improvement programme (project No. 8.1.01.2017)
 and the Saint Petersburg State University grant for scientific investigation (No. 15.61.202.2015).}

\authorcontributions{M.B.R., A.S., M.M.O, V.N.G. and A.A contribute with all the theoretical calculations, analysis of the data and dicussion. 
S. S., C. N., L. P., C. S., C. P. and P. G. contribute with all the experimental data, their analysis and interpretation.
All the authors have participated in the discussions and revision of the manuscript written by M.B.R., A.S., S.S., V.N.G. and A.A.}


\conflictsofinterest{The authors declare no conflict of interest.}



\appendix
\section{}
\subsection{Minimization procedure to obtain the self-consistent mean field equations}

The general procedure for the minimization of the free energy of the metal-TCNQ model outlined in Section \ref{modelH}
is described here. Given an interacting Hamiltonian $H$ and a variational Hamiltonian $H_0$, the Bogoliubov upper bound
for the free energy reads

\begin{eqnarray}
F&\leq&F_0+\left\langle H-H_0 \right\rangle_0,
\label{eqBgineq}
\end{eqnarray}
where, by definition,

\begin{eqnarray}
&&F=-T\ln Z
\quad\quad\mbox{and}\quad\quad
F_0=-T\ln Z_0,\nonumber\\
&&Z=\Tr\left\{e^{-\beta H}\right\}
\quad\mbox{and}\quad
Z_0=\Tr\left\{e^{-\beta H_0}\right\},\nonumber\\
&&\left\langle A\right\rangle_0=\frac{1}{Z_0}\Tr\left\{A e^{-\beta H_0}\right\},\quad\quad\forall A.
\end{eqnarray}

Using the mean-field Hamiltonian of Eq.~(\ref{eqHlochi123}) 
in the place of $H_0$, we obtain the upper bound for the free energy, which needs  subsequently to be minimized 
with respect to the order parameters $\left\langle\bm{S}_i\right\rangle$.

To analyze the XMCD data taken at normal incidence, we consider a magnetic field applied along the $Z$-axis,
$\bm{B}=(0,0,B)$.
The paramagnetic partition function of a single isolated spin reads

\begin{equation}
Z_1(h_z)=\sum_{S_z=-S}^{S}e^{-\beta\left(h_zS_z+DS_z^2\right)}.
\end{equation}

The corresponding spin average value can be found in this case by differentiating the free energy
$\left\langle\bm{S}\right\rangle_0=\partial F_1/\partial \bm{h}$, where $F_1=-T\ln Z_1$.

Two order parameters $S_a$ and $S_b$ are needed to describe the (checkerboard) antiferromagnetic phase.
They represent the $z$-components of the spins depicted in Fig.~\ref{figureMnTCNQsuperexchange}\,(a).
The upper bound to the free energy reads (per unit cell):

\begin{eqnarray}
F_{\textrm{AF}}&=&\left(J_x+J_y\right)S_aS_b+J_d\left(S_a^2+S_b^2\right)+\frac{1}{2}F_1(h_a)+\frac{1}{2}F_1(h_b)\nonumber\\
&&-(J_x+J_y)\left(S_a-\frac{\partial F_1(h_a)}{\partial h_a}\right)\left(S_b-\frac{\partial F_1(h_b)}{\partial h_b}\right)\nonumber\\
&&-J_d\left(S_a-\frac{\partial F_1(h_a)}{\partial h_a}\right)^2
-J_d\left(S_b-\frac{\partial F_1(h_b)}{\partial h_b}\right)^2,\nonumber\\
\label{eqFreeAF}
\end{eqnarray}

with

\begin{eqnarray}
h_a&=&\textsl{g}\mu_BB-2(J_x+J_y)S_b-4J_dS_a,\nonumber\\
h_b&=&\textsl{g}\mu_BB-2(J_x+J_y)S_a-4J_dS_b.
\label{eqhahbAF}
\end{eqnarray}

The terms in the last two lines of Eq.~(\ref{eqFreeAF}) come from the average $\left\langle H-H_0 \right\rangle_0$ in Eq.~(\ref{eqBgineq}).
These terms are required only when looking for the global minimum of  $F_{\textrm{AF}}(S_a,S_b)$,
which is carried out over the domain $-S\leq S_a<S$ and $-S\leq S_b<S$.
The values of $S_a$ and $S_b$ at the global minimum correspond then to the mean-field solution.
Alternatively, we can use the stationarity condition 

\begin{equation} 
\partial F_{\textrm{AF}}/\partial S_a=0\quad\quad\mbox{and}\quad\quad \partial F_{\textrm{AF}}/\partial S_b=0,
\label{eqSaSbstationaryF}
\end{equation}

to obtain the two coupled equations ~(\ref{eqSaSbselfconsistAF}).

These self-consistency equations of the mean-field theory
need to be solved for $S_a$ and $S_b$ by 
substitution of  the expressions for $h_a$ and $h_b$ from Eq.~(\ref{eqhahbAF}). 
Since several solutions can be found, we use the condition of least free energy value 
to select the physical solution.
In practice, it is convenient to find a rough approximation for $S_a$ and $S_b$ by looking for the global minimum of $F_{\textrm{AF}}(S_a,S_b)$
on a discrete grid and then refine the obtained solution by iteratively substituting it into the self-consistency equations~(\ref{eqSaSbselfconsistAF}).


\reftitle{References}





\end{document}